\documentclass{ws-procs9x6}

\begin{document}

\title{Using Quantum entanglement to study CP and  CPT violations }

\author{Yu Shi}

\address{ 
Department of Physics, Fudan University\\ Shanghai, 200433, China\\
$^*$E-mail: yushi@fudan.edu.cn}

\begin{abstract}

We report some general phenomenological results concerning CP and CPT violations in joint decays of entangled pseudoscalar neutral mesons~\cite{shi1,huang1,shi2,huang2}.

\end{abstract}

\bodymatter

\section{Introduction}

Quantum entanglement refers to the situation that the  quantum state of a composite system is not a direct product of its subsystems. It significance was  discovered by Einstein, Podolsky and Rosen~\cite{epr}.    Sch\"{o}dinger coined the term and regarded it as  ``the characteristic trait of quantum mechanics, the one that enforces its entire departure from classical lines of thought.''~\cite{sch}.
Pseudoscalar neutral mesons are copiously generated as  pairs  entangled in the flavor space~\cite{kloe,babar,belle,aquines,louvot,soni,akerib}.    CP and CPT violating parameters can be measured by studying the joint decays of entangled  meson pairs~\cite{kloe,early,kos2,soni,shi1,huang1,shi2,huang2}. This provides a venue of searching standard model extension~\cite{kos2}.

\section{Single meson states}

$|M^0\rangle$  and $|\bar{M}^0\rangle$ are  eigenstates
of parity $P$ both with eigenvalue $-1$, and of a characteristic flavor  ${\cal F}$ with eigenvalues $\pm 1$.
$C|M^0\rangle = -|\bar{M}^0\rangle$, $C|\bar{M}^0\rangle = -|M^0\rangle$. Hence the eigenstates of CP are
$
|M_\pm \rangle =
\frac{1}{\sqrt{2}}(|M^0\rangle \pm  |\bar{M}^0\rangle), $
with eigenvalues $\pm 1$.
In the Wigner-Weisskopf approximation, the weak decay of a  meson can be described by  
$
i\frac{\partial}{\partial t}|M(t)\rangle = H |M(t)\rangle,
$
where
$
H=\left(
\begin{array}{cc}
H_{00} & H_{0\bar{0}} \\
H_{\bar{0}0} & H_{\bar{0}\bar{0}}
\end{array}\right)$. It is defined that $
\frac{1-\epsilon_M}{1+\epsilon_M}  \equiv  \sqrt{ \frac{H_{\bar{0}0}}{ H_{0\bar{0}}} } \equiv \frac{q}{p}$, $\delta_M  \equiv \frac{ H_{\bar{0}\bar{0}} -H_{00}}{ \sqrt{ H_{0\bar{0}}H_{\bar{0}0} } }$.
\begin{itemize}
\item
If CP or T is conserved indirectly, then $\epsilon_M =0$.
\item
If CPT or CP is conserved indirectly, then $\delta_M =0$.
\end{itemize}

The eigenstates $|M_S\rangle$ and $|M_L\rangle$ of $H$, corresponding to the eigenvalues $\lambda_S$ and $\lambda_L$ respectively, are found by diagonalizing $H$.  Starting as  $|M_S\rangle$, the state of a single meson  evolves as $|M_S(t) \rangle = e^{-i\lambda_S t} |M_S\rangle$. Starting as $|M_L\rangle$, the state evolves as $|M_L(t) \rangle  =  e^{-i\lambda_L t} |M_L\rangle$. Based on this, one obtains $|M^0(t) \rangle$, which starts as $|M^0\rangle$; $|\bar{M}^0(t)\rangle$, which starts as  $|\bar{M}^0\rangle$; $|M_{+}(t)\rangle$, which starts as $|M_+\rangle$; and  $|M_{-}(t)\rangle$, which starts as $|M_-\rangle$.

To characterize direct violations, for decays into flavor eigenstates $|l^{\pm}\rangle$ with eigenvalue $\pm 1$, we define decay amplitudes $R^+ \equiv \langle l^+|{\cal H}|M^0\rangle$, $S^+ \equiv \langle l^+|{\cal H}|\bar{M}^0\rangle$,  $S^- \equiv \langle l^-|{\cal H}|M^0\rangle$, $R^- \equiv \langle l^-|{\cal H}|\bar{M}^0\rangle$. They can be related to quantities $a,b,c,d$ usually defined~\cite{kloe,shi2}.
\begin{itemize}
\item
If CP is conserved directly, then   $R^+=R^-$ and $S^+=S^-$.
\item If CPT is conserved directly, then $(R^+)^*=R^-$ and $(S^+)^*=S^-$.
\item If $\Delta {\cal F} = \Delta Q$ rule is respected, then we have $S^\pm =0$.
\end{itemize}

For decays into CP eigenstates   $|h^{\pm}\rangle$ with eigenvalue $\pm 1$, we define decay amplitudes
$ Q^+  \equiv \langle h^+|{\cal H} |M_{+}\rangle,$
$ X^{+}  \equiv \langle h^-|{\cal H} |M_{+}\rangle,$
$ X^-  \equiv \langle h^+|{\cal H} |M_{-}\rangle,$
$ Q^-  \equiv \langle h^-|{\cal H} |M_{-}\rangle. $ These newly defined quantities are convenient.
\begin{itemize}
\item
If CP is conserved directly, then $X^{\pm}=0$.
\item If CPT is conserved directly, then $X^{\pm}$ is purely imaginary.
\end{itemize}

\section{Entangled states}

The $C= - 1$ entangled state of a pair of pseudoscalar mesons is $
|\Psi_-\rangle   =   \frac{1}{\sqrt{2}}(|M^0\rangle_a|\bar{M}^0\rangle_b
-|\bar{M}^0\rangle_a|M^0\rangle_b) =  \frac{1}{\sqrt{2}}(|M_-\rangle_a|M_+\rangle_b-
|M_+\rangle_a|M_-\rangle_b), $
which is  produced for kaons at $\phi$ resonance~\cite{kloe},  and for B mesons by $\Upsilon (4s)$ resonance~\cite{babar,belle} and by $\Upsilon$(5S) resonance with a large branch ratio~\cite{aquines,louvot,soni}.
The $C= + 1$ entangled state   is $
|\Psi_+\rangle =    \frac{1}{\sqrt{2}}(|M_0\rangle|\bar{M}_0\rangle+|\bar{M}_0\rangle|M_0\rangle)
 = \frac{1}{\sqrt{2}}(|M_+\rangle_a|M_+\rangle_b -
|M_-\rangle_a|M_-\rangle_b),$
which  is produced for B mesons by    $\Upsilon$(5S) resonance  with  some branch ratio~\cite{aquines,louvot,soni} and above  $\Upsilon (4s)$ resonance~\cite{akerib}.

Although physically a single meson cannot be in a CP eigenstate $|M_\pm\rangle$ because of CP violation, the entangled states can be {\em exactly} written in terms of  $|M_\pm\rangle$, this makes the expression $|M_\pm(t)\rangle$ and the decay amplitudes $Q^\pm$ and $X^\pm$ meaningful and useful.

Starting from  $|\Psi_-\rangle$,  the entangled meson pair decay to certain products at $t_a$ and $t_b$, hence the time-dependent state is
$
|\Psi_-(t_a,t_b)\rangle = \frac{1}{\sqrt{2}}(|M^0(t_a)\rangle_a|\bar{M}^0(t_b)\rangle_b
-|\bar{M}^0(t_a)\rangle_a|M^0(t_b)\rangle_b)
 = \frac{1}{\sqrt{2}}(|M_-(t_a)\rangle_a|M_+(t_b)\rangle_b
-|M_+(t_a)\rangle_a|M_-(t_b)\rangle_b).$
Similarly, starting as $|\Psi_+\rangle$, $
|\Psi_+(t_a,t_b)\rangle = \frac{1}{\sqrt{2}}(|M^0(t_a)\rangle_a|\bar{M}^0(t_b)\rangle_b
+|\bar{M}^0(t_a)\rangle_a|M^0(t_b)\rangle_b)
 = \frac{1}{\sqrt{2}}(|M_+(t_a)\rangle_a|M_+(t_b)\rangle_b
-|M_-(t_a)\rangle_a|M_-(t_b)\rangle_b).$

\section{Joint decay rates}

For state $|\Psi(t_a,t_b)\rangle$, the joint  rate that particle a decays to $f$ at $t_a$ while particle b decays to $g$ at $t_b$ is
$
I (f,t_a;g,t_b) = |\langle f,g|{\cal H}_a {\cal H}_b |\Psi(t_a,t_b)\rangle|^2,$
where ${\cal H}_a$ and ${\cal H}_b$ represent the Hamiltonians governing the weak decays of $a$ and $b$, respectively.  In experiments, it is more convenient to use the integrated rate
$
I(f,g,\Delta t) = \int_0^\infty I(f, t_a; g, t_a+\Delta t) dt_a.$
Then one can find the asymmetry between the joint decays to $f$ and $g$ and the joint decays to $f'$ and $g'$,
$
A(fg, f'g', \Delta t) \equiv \frac{I[f,t_a; g, t_a+\Delta t ]-I[f',t_a; g', t_a+\Delta t ]}{I[f,t_a; g, t_a+\Delta t ]+ I[f',t_a; g', t_a+\Delta t ]}
  =   \frac{I[f, g, \Delta t]-I[f', g', \Delta t]}
{I[f, g, \Delta t]+I[f', g', \Delta t]}.$

We have considered the following cases of the final states: (1) the decay products are  flavor eigenstates $|l^\pm\rangle$, with the equal-flavor asymmetry $A(l^+l^+,l^-l^-,\Delta t)$ and the unequal-flavor asymmetry $A(l^+l^-,l^-l^+,\Delta t)$; (2) the decay products are  CP eigenstates $|h^\pm\rangle$, with the  equal-CP asymmetry $A(h^+h^+,h^-h^-,\Delta t)$ and the unequal-CP asymmetry    $A(h^+h^-,h^-h^+,\Delta t)$; (3)  the decay products $|h_1\rangle$ and $|h_2\rangle$ are CP conjugates.

\section{General results on joint decays of $|\Psi_-\rangle$~\cite{shi2}  } 

{\bf Theorem 1}
   If the equal-flavor asymmetry is nonzero,  then  there exists one or two of the following violations: (1) CP is violated indirectly, (2) both CP and CPT are violated directly.

{\bf Theorem 2}  If the equal-flavor asymmetry  is nonzero while CPT is assumed to be conserved both directly and indirectly, then in addition to indirect CP violation, we can draw the following conclusions: (1) $|q/p| \neq 1$, i.e.  T must also be  violated indirectly; (2)    $|\langle l^+ |{\cal H}|\bar{M}_0\rangle| \neq |\langle l^+|{\cal H}  |M^0\rangle|$,  $|\langle  l^-|{\cal H}  |M^0\rangle | \neq |\langle l^- |{\cal H}|\bar{M}_0\rangle| $, despite   $ \langle l^+|{\cal H}  |M^0\rangle = \langle l^- |{\cal H}|\bar{M}_0\rangle^*$ and $\langle l^+ |{\cal H}|\bar{M}_0\rangle  = \langle  l^-|{\cal H}  |M^0\rangle^* $.

{\bf Theorem 3}  If  the unequal-flavor asymmetry is nonzero, then CP must be violated,  directly or indirectly or both.

{\bf Theorem 4} If   the unequal-flavor asymmetry is nonzero for $\Delta t \neq 0$ while CPT is assumed to be conserved both directly and indirectly, then  we can draw the following conclusions: (1)   $|\langle l^+ |{\cal H}|\bar{M}_0\rangle|= |\langle  l^-|{\cal H}  |M^0\rangle | \neq |\langle l^+|{\cal H}  |M^0\rangle|= |\langle l^- |{\cal H}|\bar{M}_0\rangle| $; (2)  $\langle l^{-}|{\cal H} |M^0\rangle = \langle l^{+}|{\cal H}|\bar{M}^0\rangle^* \neq 0$, which means $\Delta {\cal F} =\Delta Q$ rule must be violated.

{\bf Theorem 5}  The equal-CP asymmetry  is always a constant independent of $\Delta t$.

{\bf Theorem 6}   For $\Delta t =0$, the unequal-CP asymmetry  vanishes,   no matter whether CP or CPT is violated.

{\bf Theorem 7}  If any equal-CP joint decay rate is nonzero, then CP must be violated,  directly or indirectly or both.

\section{General results on joint decays of $|\Psi_+\rangle$~\cite{huang2}} 

{\bf Theorem 8: } If the unequal-flavor asymmetry  is nonzero, then CP must be violated indirectly.

{\bf Theorem 9: } If the $I(l^+,l^-,\Delta t)$ and $I(l^-,l^+,\Delta t)$ depend on the first order of $\epsilon_M$, then CP must also be violated directly.

{\bf Theorem 10:}  If the unequal-flavor asymmetry  depends on the first or second order of $\epsilon_M$, then  $\Delta {\cal F}=\Delta Q$ rule is violated.

{\bf Theorem 11:} The deviation of the unequal-CP joint decay rate $I(h^+,t_a;h^-,t_b)$ or $I(h^-,t_a;h^+,t_b)$ from zero  implies direct CP violation.

{\bf Theorem 12:} Suppose  $|h_1\rangle$ and $|h_2\rangle$ are CP conjugates.  If $I(h_1,h_2;\Delta t)$ and $I(h_2,h_1;\Delta t)$  depend on the first order   of   $\epsilon_M$, then CP is violated directly.

In addition, we have derived various quantitative relations of  the indirect violating parameters with the decay   asymmetries of $|\Psi_-\rangle$~\cite{shi1,shi2}, with those of $|\Psi_+\rangle$~\cite{huang2} and with four asymmetries defined for some time-ordered integrated rates of $|\Psi_-\rangle$ and  $|\Psi_+\rangle$~\cite{huang1}.

This work is supported by the National Science Foundation of China (Grant No. 10875028).


\begin{thebibliography}{xx}
\bibitem{shi1} Y. Shi, Euro. Phys. J. C {\bf 72}, 1907(2012).
\bibitem{shi2} Y. Shi, arXiv:1306.2676.
\bibitem{huang2} Z. Huang and Y. Shi,  arXiv:1307.4459.
\bibitem{huang1} Z. Huang and Y. Shi, Euro. Phys. J. C {\bf 72}, 1900(2012).
\bibitem{epr} E. Einstein, B. Podolsky and N. Rosen, Phys. Rev. {\bf 47},
777 (1935).
\bibitem{sch} E. Schr\"{o}dinger,    Proc. Camb.  Phi.  Soc. {\bf 31}, 555 (1936).
\bibitem{kloe} A. Di Domenico (KLOE Collaboration), Found. Phys.
{\bf 40}, 852 (2010).
\bibitem{babar} B. Aubert {\it et al.} (BABAR Collaboration), Phys. Rev. Lett. {\bf 88}, 221802 (2002).
\bibitem{belle} A. Go  {\it et al.} (BELLE Collaboration), Phys. Rev. Lett. 99, 131802 (2007).
\bibitem{aquines} O. Aquines {\it et al.} (CLEO Collaboration), Phys. Rev. Lett. 96, 152001  (2006).
\bibitem{louvot} R. Louvot  {\it et al.} (BELLE Collaboration), Phys. Rev. Lett. 102, 021801  (2009).
\bibitem{soni} D. Atwood and A. Soni, Phys. Rev. D {\bf 82}, 036003 (2010).
\bibitem{akerib}  D. S. Skerib  {\it et al.} (CLEO II Collaboration), Phys. Rev. Lett. 67, 1692   (1991).
\bibitem{early}  J. Bernab\'{e}u, F. J. Botella and J. Rold\'{a}n, Phys. Lett. B {\bf 211}, 226 (1980);  I. Dunietz, J. Hauser and J. L. Rosner, Phys. Rev. D {\bf 35}, 2166 (1987); C. D. Buchanan {\it et al.}, Phys. Rev. D {\bf 45}, 4088 (1992); G. D'Ambrosio, G. Isidori and A. Pugliese,  in L. Maiani, G. Pancheri and N. Paver. (eds.), {\em The Second DA$\Phi$NE Physics Handbook},  (SIS-Publicazioni, Frascati, 1995).
\bibitem{kos2}  V. A. Kosteleck\'{y}, Phys. Rev. Lett. {\bf 80}, 1818 (1998); V. A. Kosteleck\'{y}, Phys. Rev. D {\bf 61}, 016002 (1999); V. A. Kosteleck\'{y}, Phys. Rev. D {\bf 64}, 076001 (2001).
\end{thebibliography}
\end{document}